# Comparative Statics in Multicriteria Search Models

Veli Safak[1][2]

## Abstract

McCall (1970) examines the search behaviour of an infinitely-lived and risk-neutral job seeker maximizing her lifetime earnings by accepting or rejecting real-valued scalar wage offers. In practice, job offers have multiple attributes, and job seekers solve a multicriteria search problem. This paper presents a multicriteria search model and new comparative statics results.

### Introduction

In the search literature, the seminal work by McCall (1970) proposes a search model in which an infinitely-lived and risk-neutral job seeker receives random real-valued wage offers from a known wage distribution. After receiving an offer, she has two options: accept the offer and end the search, or reject the offer and receive another offer in the next period. Her goal is to maximize her lifetime utility. In this framework, the optimal policy admits a threshold form, i.e., the job seeker accepts all offers above a reservation wage and rejects the others.

In this paper, I show that McCall's (1970) tractable result holds when job offers are characterized by multiple attributes and aggregated via a utility function. I also identify four distributional changes increasing the reservation utility to accept an offer.

### 1. Multicriteria Search Model

Consider a labor market in which every job offer has K attributes, i.e., $w \in \mathbb{R}^K$. Imagine an infinitely-lived job seeker who has time-separable preferences between offers. Suppose that utility function U represents the preference of each agent: $\mathbb{R}^K \to \mathbb{R}$. In each period of unemployment, the job seeker receives a random wage offer from distribution $F: \mathbb{R}^K \to [0,1]$. If she accepts the offer, then she works at the accepted job for all future periods. If she rejects the offer, then she receives an exogenously specified unemployment flow utility $\gamma > 0$ for a period and waits until the next period to receive another job offer. She discounts future utility with discount parameter $\beta \in (0,1)$. Letting $V_F(w)$ denote the maximum possible discounted lifetime utility of the job seeker with current job offer w, and the Bellman equation can be specified as follows:

$$V_F(w) = \max\{(1-\beta)^{-1} U(w), \gamma + E_F\{V_F(w')\}\} \quad (1)$$

Note that if the job seeker accepts an offer yielding to utility level $u \in \mathbb{R}_+$, she also accepts offers yielding to higher utility levels. Similarly, if the job seeker rejects an offer yielding to utility level $u' \in \mathbb{R}_+$, then she also rejects offers yielding to lower utility levels. Since $\mathbb{R}_+$ is a convex and ordered set, there exists a unique level of utility for which the job seeker is indifferent between accepting and rejecting it. Altogether, the following value function obtains:

$$V_F(w) = \begin{cases} \dfrac{U(w)}{1-\beta} & U(w) \geq u_F \\ \dfrac{u_F}{1-\beta} = \gamma + \beta E_F\{V_F(w')\} & U(w) \leq u_F \end{cases} \quad (2)$$

Reservation utility $u_F$ solves the following equality:

$$u_F = \gamma + \frac{\beta}{1-\beta} E_F\{\max\{U(w') - u_F, 0\}\} \quad (3)$$

**Theorem 1** Consider a set of real-valued utility functions **U** on $\mathbb{R}^K$ such that $H \in \mathbf{U}$ implies that $(1-\beta)^{-1}\max\{H, u\} \in \mathbf{U}$ for all $u \in \mathbb{R}_+$. Furthermore, let $E_F\{H(w')\} \geq E_G\{H(w')\}$ for all $H \in \mathbf{U}$. Under these assumptions, the reservation utility is higher under F relative to G.

**Proof of Theorem 1** According to Equation (2), $V_F(w)$ can be written as follows for the optimal $u_F \in \mathbb{R}_+$:

$$V_F(w) = (1-\beta)^{-1} \max\{U(w), u_F\} \quad (4)$$

We have $V_G(w) \leq V_F(w)$ as $E_F\{H(w')\} \geq E_G\{H(w')\}$ for all $H \in \mathbf{U}$ and $(1-\beta)^{-1}\max\{H, u\} \in \mathbf{U}$:

$$V_G(w) = (1-\beta)^{-1} \max\{U(w), \gamma + \beta E_G\{\max\{U(w'), u_G\}\}\}$$
$$\leq (1-\beta)^{-1} \max\{U(w), \gamma + \beta E_F\{\max\{U(w'), u_G\}\}\}$$
$$\leq (1-\beta)^{-1} \max\{U(w), \gamma + \beta E_F\{\max\{U(w'), u_F\}\}\}$$
$$= V_F(w).$$

---

[1] Department of Business Administration, Carnegie Mellon University Qatar, E-mail: vsafak@andrew.cmu.edu

[2] I thank Axel Z. Anderson for his invaluable guidance. I would also like to thank James W. Albrecht and Luca Anderlini for their constructive comments. I would further like to thank Roger Lagunoff, John Rust, Laurent Bouton, Dan Cao, and Yusufcan Masatlioglu for helpful discussions.

The first inequality follows from the assumptions of the theorem. The second inequality is a consequence of the optimality of $u_F$ under distribution F. As a result, the continuation value of search is higher under F relative to G:

$$E_G\{V_G(w')\} \leq E_F\{V_G(w')\} \leq E_F\{V_F(w')\}.$$

Consequently, the reservation utility is higher under F relative to G. ∎

The critical observation is that the value function in Equation (4) is a positive affine transformation[3] of the truncation[4] of a positive affine transformation of the utility function. Suppose that utility function U(w) of w belongs to set of functions **U** which is closed under positive affine transformation and truncation. For such sets of functions, Theorem 1 can be used to obtain new comparative static results and understand the relationship between optimal strategy and the distribution of job offers. In the next section, I show that some standard sets of functions are closed under positive affine transformation and truncation.

## 2. Multicriteria Comparative Statics

In the search framework presented by McCall (1970), where $w \in \mathbb{R}$ and $U(w) = w$, there are two well-known comparative static results. If (a) the probability of receiving a wage offer at least as high as $w \in \mathbb{R}$ is higher under wage distribution F relative to G, or (b) F is obtained from G by spreading out probability mass without changing the mean, then the reservation wage is higher under F compared to G.

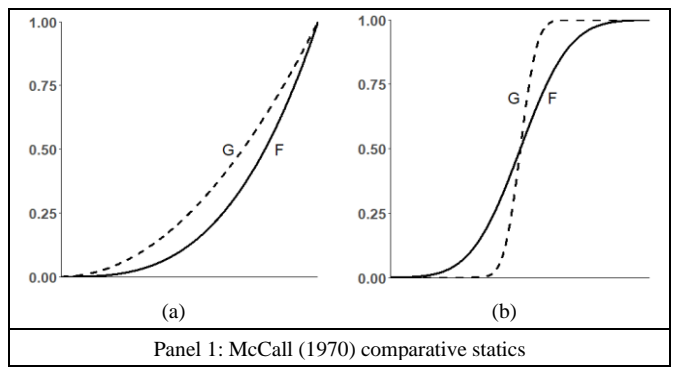

Panel 1: McCall (1970) comparative statics

First, I show that McCall's (1970) comparative static results apply to the multicriteria search models.

**Definition 1** Utility function $U: \mathbb{R}^K \to \mathbb{R}$ is called

(a) convex if $U(\lambda w + (1-\lambda)w') \leq \lambda U(w) + (1-\lambda)U(w')$ for all $w, w' \in \mathbb{R}^K$ and $\lambda \in [0,1]$;

(b) componentwise convex if for all $w, w' \in \mathbb{R}^K$ such that $w_j = w'_j$ for all $j \neq i$ for some $i \in \{1, \ldots, K\}$ and $\lambda \in [0,1]$, it holds that $U(\lambda w + (1-\lambda)w') \leq \lambda U(w) + (1-\lambda)U(w')$.

**Definition 2** Given set of functions **U**, offer distribution $F: \mathbb{R}^K \to [0,1]$ dominates offer distribution $G: \mathbb{R}^K \to [0,1]$ on set **U** if $E_F\{U(w')\} \geq E_G\{U(w')\}$.

**Theorem 2** The reservation utility is higher under F compared to G if for one of the following sets (*) F dominates G on the set and (**) the utility function belongs to the set:
(a) set of increasing functions,
(b) set of convex functions, and
(c) set of component-wise convex functions.

Theorem 2 shows that McCall's (1970) classic comparative static results apply to a broader set of utility functions in the multicriteria framework (for proof, see the Appendix).

In the multicriteria framework, we can also explore how complementarity and dependence between offer attributes affect the job seeker's decision. In a practical sense, one might expect to observe complementarities between offer attributes. For example, earning a high wage in a city with a wide array of entertainment could be better for a job seeker relative to earning a high wage in the middle of nowhere. Furthermore, we can also expect dependence between offer attributes. High wage offers are often associated with better social benefits and health insurance packages.

In the matching literature, complementarity is often modelled by using supermodularity.

**Definition 3** Utility function $U: \mathbb{R}^K \to \mathbb{R}$ is supermodular if for all $w, w' \in \mathbb{R}^K$, it holds that

$$U(w \wedge w') + U(w \vee w') \geq U(w) + U(w')$$

where $\wedge$ and $\vee$ represent the component-wise minimum and maximum.

It is important to note that the set of supermodular functions is not closed under truncation.

**Example 1** Let $U(1,1) = U(2,2) = 5$, $U(1,2) = -5$, and $U(2,1) = 14$. Note that U is supermodular.

$$U(1,1) + U(2,2) = 10 \geq 9 = U(1,2) + U(2,1)$$

On the other hand, the truncation of U(w) of w, defined as $\max\{U(w), 0\}$ and denoted by $T\{U(w)\}$ is not supermodular.

$$U(1,1) + U(2,2) = 10 < 14 = U(1,2) + U(2,1)$$

In the appendix, I show that the set of increasing supermodular functions, i.e., both increasing and supermodular, is closed under truncation and positive affine transformation. Thus, the following theorem obtains.

**Theorem 3** The reservation utility is higher under F relative to G if F dominates G on the set of increasing supermodular functions, and the utility function is increasing supermodular.

Theorem 3 states that if the job seeker's preference exhibits complementarities across offer attributes, then an increase in statistical dependence between offer attributes increases the reservation utility and contracts the set of acceptable offers.

---
[3] The positive affine transformation of U(w) with parameters $m \in \mathbb{R}_+$ and $n \in \mathbb{R}_+$ is specified as $mU(w) + n$
[4] The truncation of U(w) is $\max\{U(w), 0\}$.

Muller and Scarsini (2012) propose ultramodularity as a multivariate convexity notion. A function is *ultramodular* if it is both supermodular and componentwise convex. The next theorem shows that ultramodularity can be used in multicriteria search models as a generalization of convexity while maintaining tractable comparative static results.

**Theorem 4** The reservation utility is higher under F relative to G if F dominates G on the set of increasing ultramodular functions, and the utility function is increasing ultramodular.

### Conclusion

This paper presents a simple framework for multicriteria search models. It shows that the optimal strategy admits a threshold policy. Based on this form, it also proves that the classic comparative static results of single criteria search models apply (Theorem 2). It also offers new comparative static results with no counterpart in single criteria search models and establishes the relationship between increasing dependence between offer attributes and optimal search policy for a broad set of utility functions with minimal functional restrictions.

### References

[1] Chade, H., Eeckhout, J., Smith, L. Sorting through search and matching models. *Journal of Economic Literature* **55(2)**, 1-52 (2017).

[2] Lim, C., Bearden, J. N., Smith J.C. Sequential search with multi-attribute options. *Decision Analysis* **3(1)**, 3-15 (2006).

[3] MacQueen, J.B. Optimal policies for a class of search and evaluation problems. *Management Science* **10(4)**, 746-759 (1964).

[4] Marinacci, M., Montrucchio, L. Ultramodular functions. *Math. Oper. Res.* **30**, 311-332 (2005).

[5] McCall, J.J. Economics of information and job search. *Quarterly Journal of Economics* **84**, 113-126 (1970).

[6] Muller, A., Scarsini, M. Fear of loss, inframodularity, and transfers. *Journal of Economic Theory* **147(4)**, 1490-1500 (2012).

[7] Weitzman, M. L. Optimal search for the best alternative. *Econometrica* **47(3)**, 641-654 (1979).

### Appendix

**Proof of Theorem 2.** It suffices to show the following sets are preserved under truncation and positive affine transformation.

(a) set of increasing functions: trivial

(b) set of convex functions

Let $U: \mathbb{R}^K \to \mathbb{R}$ be a convex function. For all $w, w' \in \mathbb{R}^K$ and $\lambda \in [0,1]$, it holds that

$$U(\lambda w + (1-\lambda)w') \leq \lambda U(w) + (1-\lambda)U(w').$$

Truncation:

$T\{U(\lambda w + (1-\lambda)w')\} = \max\{U(\lambda w + (1-\lambda)w'), 0\}$

$\leq \max\{\lambda U(w) + (1-\lambda)U(w'), 0\}$   (due to convexity of U)

$\leq \lambda \max\{U(w), 0\} + (1-\lambda)\max\{U(w'), 0\}$

$\leq \lambda T\{U(w)\} + (1-\lambda)T\{U(w')\}$

Positive affine transformation with parameters m and n:

$m\{U(\lambda w + (1-\lambda)w')\} + n$

$\leq m\{\lambda U(w) + (1-\lambda)U(w')\} + n$   (due to convexity of U)

$\leq \lambda\{mU(w) + n\} + (1-\lambda)\{mU(w') + n\}$

(c) set of componentwise convex functions: it can be proved by following the logic in the previous proof. ∎

**Proof of Theorem 3.**

It suffices to show that the set of increasing supermodular functions is closed under truncation and affine transformation. Let $U: \mathbb{R}^K \to \mathbb{R}$ be an increasing supermodular function. For all $w, w' \in \mathbb{R}^K$, it holds that

$$U(w \wedge w') + U(w \vee w') \geq U(w) + U(w').$$

The closedness under positive affine transformation is trivial.

To prove the closedness under truncation define $S(w, w')$ as follows:

$$T\{U(w \wedge w')\} + T\{U(w \vee w')\} - T\{U(w)\} - T\{U(w')\}.$$

It suffices to show that $S(w, w') \geq 0$. Note that the value of $T\{U(\cdot)\}$ depends on the sign of $U(\cdot)$. Furthermore, many sign combinations for $U(w \wedge w'), U(w \vee w'), U(w)$, and $U(w')$ are not available due to $U(\cdot)$ being an increasing function. After eliminating these cases, the number of potential combinations drops from 16 to 6.

C1: $U(w \vee w') < 0, U(w \wedge w') < 0, U(w) < 0, U(w') < 0$

$S(w, w') = 0$

C2: $U(w \vee w') \geq 0, U(w \wedge w') < 0, U(w) < 0, U(w') < 0$

$S(w, w') = U(w \vee w') \geq 0$

C3: $U(w \vee w') \geq 0, U(w \wedge w') < 0, U(w) \geq 0, U(w') < 0$

$S(w, w') = U(w \vee w') - U(w) \geq 0$

C4: $U(w \vee w') \geq 0, U(w \wedge w') < 0, U(w) < 0, U(w') \geq 0$

$S(w, w') = U(w \vee w') - U(w') \geq 0$

C5: $U(w \vee w') \geq 0, U(w \wedge w') < 0, U(w) \geq 0, U(w') \geq 0$

$S(w, w') = U(w \vee w') - U(w) - U(w')$

$\geq U(w \vee w') + U(w \wedge w') - U(w) - U(w') \geq 0$

C6: $U(w \vee w') \geq 0, U(w \wedge w') \geq 0, U(w) \geq 0, U(w') \geq 0$

$S(w, w') = U(w \vee w') + U(w \wedge w') - U(w) - U(w') \geq 0$ ∎